\documentclass[12pt]{article}
\usepackage{amsfonts}
\usepackage{amssymb}
\usepackage{amsmath}
\usepackage{graphicx}
\usepackage{amsthm}

\theoremstyle{plain}

\newtheorem*{conjecture*}{Conjecture}

\newcommand{\expect}[1]{
 \left\langle #1\right\rangle
}

\def\tr{\mbox{tr}}

\def\minor{\mbox{minor}}

\def\be{\begin{equation}}
\def\ee{\end{equation}}

\begin{document}

\begin{titlepage}

\begin{flushright}
hep-th/0505029\\
NSF-KITP-05-27
\end{flushright}

\vspace{5mm}

\begin{center}
{\huge  Supersymmetry Breaking from a Calabi-Yau Singularity}\\
\vspace{3mm}
\vspace{1mm}
\end{center}

\vspace{1mm}
\begin{center}
{\large D.~Berenstein$^*$, C.~P.~Herzog$^\dag$, 
P.~Ouyang$^*$, and S.~Pinansky$^*$
}\\
\parbox{3in}{ \begin{center}
$\dag$ Kavli Institute for Theoretical Physics\\
University of California \\ Santa Barbara, CA  93106, USA\\
\end{center}}
\parbox{3in}{ \begin{center} $*$ Physics Department\\
University of California \\ Santa Barbara, CA  93106, USA\\
\end{center}}\\

\end{center}

\vspace{3mm}

\begin{abstract}

\noindent
We conjecture a geometric criterion for determining whether supersymmetry
is spontaneously broken in certain string backgrounds.
These backgrounds
contain wrapped branes at Calabi-Yau singularites with obstructions to
deformation of the complex structure.  We motivate our conjecture with
a particular example: the $Y^{2,1}$ quiver gauge theory corresponding
to a cone over the first del Pezzo surface, $dP_1$.  This setup can be
analyzed using ordinary supersymmetric field theory methods, where we
find that gaugino condensation drives a deformation of the chiral ring
which has no solutions. We expect this breaking to be a general
feature of any theory of branes at a singularity with a smaller number
of possible deformations than independent anomaly-free fractional
branes.  
\vfil
\end{abstract}
\begin{flushleft}
May 2005
\end{flushleft}
\vfil
\end{titlepage}
\newpage

\section{Introduction}

Supersymmetry and supersymmetry breaking are central ideas both in
contemporary particle physics and in mathematical physics.  In this
paper, we argue that for a large new class of D-brane models there
exists a simple geometric criterion which determines whether
supersymmetry breaking occurs.

The models of interest are based on Calabi-Yau singularities with
D-branes placed at or near the singularity. By taking a large volume
limit, it is possible to decouple gravity from the theory, and ignore
the Calabi-Yau geometry far from the branes.  Although one typically
begins the construction with branes free to move throughout the
Calabi-Yau space, many interesting theories include fractional branes
(or wrapped branes), which are D-branes stuck at the singularity and
which cannot move away from it. These fractional branes can lead to
topology changes in the geometry via gaugino condensation, which can
be understood in terms of deformations of the complex structure of the
Calabi-Yau singularity, and result in the replacement of branes by fluxes \cite{KS,GV}. Models of this type, Calabi-Yau geometries with fluxes, are among the ingredients used to construct viable cosmological models within string theory with a positive cosmological constant
\cite{KKLT}, and which, because of that positivity, ultimately break
supersymmetry.  With current approaches in the literature, the issue
of whether or not supersymmetry breaking is under control is
controversial.
 
It turns out that there exist some singular geometries which admit
fractional branes, but for which the associated complex structure
deformations are obstructed.  We will argue that in such setups a
deformation can still take place, but that the geometric obstruction
induces supersymmetry breaking.
\begin{conjecture*}
Given a set of ordinary and fractional branes probing a Calabi-Yau
singularity, supersymmetry breaking by obstructed geometry (SUSY-BOG)
occurs when the singularity admits fewer deformations than the number
of (anomaly free) independent fractional branes that one can add to
the system.
\end{conjecture*}
In this paper we will show how this breaking occurs in one particular
example, D-branes probing the singularity of the Calabi-Yau cone over
the first del Pezzo surface, $dP_1$. This case can be analyzed using
the gauge theory/gravity correspondence.  In field theory, the
mechanism that spontaneously breaks SUSY is simply gaugino
condensation.  In terms of the dual gravity theory and its local
Calabi-Yau geometry, the deformation obstruction causes the breaking.
The SUSY-BOG adds a new feature to the string theory landscape.

$dP_1$ belongs to a large class of singularities for which this
geometric obstruction is easy to understand.  We need the technical
concept from algebraic geometry of a singularity which is an
``incomplete intersection variety.''  These varieties are described by
$n$ equations in $m$ variables, where the dimension of the variety is
strictly greater than $m-n$. Although the $n$ equations or relations
are redundant, it is impossible to reduce the number of equations
further without changing the geometry.  In the case of an incomplete
intersection, the redundancy makes it hard to deform the equations
consistently, giving an obstruction to deformation.  These incomplete
intersections are believed to be more generic than ``complete
intersections,'' where the dimension of the variety is equal to $m-n$.

In the dual field theory, the $m$ complex variables can be
associated to generators of the chiral ring of the theory, while the
relations in the geometry can be understood as relations in the chiral
ring. We will develop this picture further to make this
set of ideas more precise. A generic property of fractional branes is that
they have gaugino condensation on their worldvolume, which translates
to non-trivial deformed relations in the chiral ring of the field
theory.  When it is possible to deform the equations of the chiral
ring consistently, the field theory realizes a supersymmetric vacuum.  
If it is impossible to deform the equations in the chiral
ring consistently, then the theory spontaneously breaks supersymmetry.

\subsection{The Conifold vs. $dP_1$}

Perhaps the prototypical example of a Calabi-Yau singularity with
fractional branes is the conifold singularity, in the setups of
Klebanov and Witten \cite{KW} (see also \cite{MP, Kehag}), and its
generalization to the warped deformed conifold geometry of Klebanov
and Strassler (KS) \cite{KS}.  The KS solution is an example of a
space with no obstruction to deformation, and in this case chiral
symmetry breaking and confinement in the strongly coupled dual gauge
theory may be identified with the deformation in a precise way.

Recall that the construction of the KS solution begins by placing a
stack of $N$ D3-branes and $M$ D5-branes at the tip of the conifold:
\be 
\sum_{i=1}^4 z_i^2 = 0 \ .
\label{coneq}
\ee 
The D5-branes are fractional.  In the supergravity solution, the
D-branes are replaced by their corresponding RR five-form and
three-form fluxes.  The three-form flux from the D5-branes leads to a
singularity unless the conifold is deformed:
\be
\sum_{i=1}^4 z_i^2 = \epsilon \ .
\label{coneqdef}
\ee
This $\epsilon$ parameter is interpreted as a gaugino condensate in
the dual theory and breaks the R-symmetry to ${\mathbb Z}_2$.
Moreover, cutting off the cone at a radius $\epsilon$ produces
confining behavior of strings in this geometry which are dual to
electric flux tubes in the gauge theory.

Naively, one could imagine that a very similar phenomenon occurs for
the cone over $dP_1$, but there is an obstruction.  A theorem by
Altmann \cite{Altmann} claims that there are no deformations of this
cone.  While the
conifold has the single defining equation (\ref{coneq}), we recall
below that the cone over $dP_1$ requires twenty algebraic relations in
${\mathbb C}^9$ which are not all independent.  It is impossible to
alter all twenty consistently by adding the analog of $\epsilon$ in
(\ref{coneqdef}).

The conformal analog of the $dP_1$ model with no D5-branes was first
proposed separately by Kehagias \cite{Kehag} and by Morrison and
Plesser \cite{MP} as an example of a generalized AdS/CFT
correspondence.  Type IIB string theory propagating in $AdS_5 \times
Y$, where $Y$ is a particular U(1) bundle over $dP_1$, is equivalent
to a superconformal ${\mathcal N}=1$ gauge theory whose field content
and superpotential were later derived in \cite{Feng} using techniques
from toric geometry (see Fig.~\ref{dP1}).

Recently, the metric on $Y$ was discovered.  Martelli and Sparks
\cite{Martelli} have shown that this U(1) bundle over $dP_1$,
$Y=Y^{2,1}$, is a particular example of the $Y^{p,q}$ Sasaki-Einstein
manifolds found in \cite{Gauntlett2, Gauntlett}.  Using the explicit
metric, impressive comparisons of the conformal anomaly coefficients
and R-charges between the AdS and the CFT sides were carried out for
$dP_1$ in \cite{Bertolini, Benvenuti}.

As was the case for the conifold, adding D5-branes to the tip of the cone
over $dP_1$ breaks conformal invariance and leads to a cascading gauge
theory \cite{Hanwalls, chaotic}.  Recently, a supergravity solution
dual to this cascading theory was derived by Herzog, Ejaz, and
Klebanov (HEK) \cite{HEK}.  The HEK solution shares many of the
features of the predecessor of the KS solution, the Klebanov-Tseytlin
(KT) solution \cite{KT}.  Like the KT solution, the HEK solution is
singular in the infrared (small radius); at large radius, in the
ultraviolet, the KT solution approaches the KS solution.  The large
radius region can be used to calculate geometrically the NSVZ beta
functions and logarithmic running in the number of colors, and this
geometric behavior matches our field theoretic understanding of a
duality cascade.

However, there currently is no known analog of the KS solution for $dP_1$ that smooths the singularity in the infrared.  The motivation behind this research was to understand the infrared physics of $dP_1$.

\subsection{Summary of the $dP_1$ Argument}

Since the Calabi-Yau deformations are obstructed, we argue that
supersymmetry is broken at the bottom of the duality cascade for
$dP_1$.  Our argument has three field theoretic legs.  First we argue
that there is no superconformal fixed point to which the cascading
theory may flow, ruling out strong coupling behavior that is not
confinement.  Next, turning on Fayet-Iliopoulos terms in the field
theory, we show that the theory should confine in the infrared.  Last,
using the Konishi anomaly \cite{Konishi}, we show there are no
solutions to the F-term equations.

The last step requires some immediate clarification.  The field theory
analysis is well suited for studying vacua where mesonic operators get
expectation values -- the mesonic branch -- and is less so for the
baryonic branch.  We will show that the mesonic branch of the $dP_1$
field theory spontaneously breaks supersymmetry.

We have a traditional field theory argument that supersymmetry is not
preserved along the baryonic branch, but the argument is not as
rigorous as the Konishi anomaly approach.  It may be that there exists
a supersymmetric, pure flux supergravity solution (which does not
involve a Calabi-Yau manifold), similar to the recent $SU(3)$
structure solutions of \cite{Butti, Grana}.  Recall that the KS
solution represents a particular point in the baryonic branch of the
conifold field theory \cite{GHK}.  As one moves along the baryonic
branch, one finds that the geometry is no longer Calabi-Yau but is
instead an $SU(3)$ structure solution where probe D3-branes break
supersymmetry.

For generic initial number of D3- and D5-branes in the UV, we expect
to have left-over D3-branes at the bottom of the cascade.  The KS
solution represents a particular point in the baryonic branch where
these D3-branes will remain supersymmetric.  However, even if there is
an $SU(3)$ structure solution for $dP_1$, the left-over D3-branes will
always break supersymmetry.

We emphasize that the supersymmetry breaking is an infrared effect.
The warp factor in the geometry will lead to an exponential separation
between the ultraviolet and infrared scales, giving a natural way of
achieving low-scale supersymmetry breaking should we embed this cone
over $dP_1$ in a compact Calabi-Yau.  An optimistic point of view is
that one might be able to build supergravity solutions from
compactified versions of this cone over $dP_1$, and using these
solutions, produce realistic cosmological models similar to
\cite{KKLT} where all stringy moduli are stabilized and supersymmetry
is broken in a controlled way.  Such solutions deserve further study.

\section{The Gauge Theory for $dP_1$}
\label{sec:dp1}

\begin{figure}
\begin{center}
\includegraphics[width=3in]{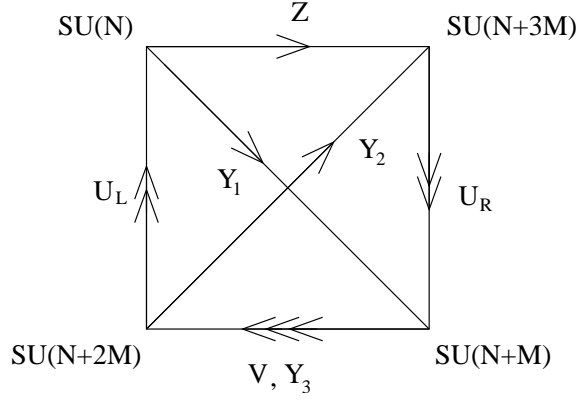}
\end{center}
\vfil
\caption{The quiver theory for $Y^{2,1}$, also known as an irregular
$U(1)$ fibration of ${\mathbb P}^2$ 
blown up at a point.}
\label{dP1}
\end{figure}

The quiver theory for the first del Pezzo has four gauge groups and a
number of bifundamental fields conveniently summarized using a quiver
diagram (see Fig.~\ref{dP1}).
The superpotential for the first del Pezzo takes the form
\be
W = \epsilon_{\alpha\beta} U_L^\alpha V^\beta Y_1 - \epsilon_{\alpha \beta}U_R^\alpha Y_2 V^\beta
- \epsilon_{\alpha\beta} U_L^\alpha Y_3 U_R^\beta Z \ .
\label{WdP1}
\ee
There is an $SU(2)$ global symmetry: 
the indices $\alpha$ and $\beta$ are either one or two and $\epsilon_{12}=1$.

In the case $M=0$, the gauge theory has a superconformal fixed point
with an AdS/CFT dual description: type IIB string theory in a $AdS_5
\times Y^{2,1}$ background.  Adding $M$ D5-branes breaks the
superconformal symmetry and starts a Seiberg duality cascade, as
argued in \cite{Hanwalls, chaotic, HEK}.  In the limit $M \ll N$, we
can calculate the NSVZ $\beta$-functions and find that generically the
$SU(N+3M)$ group runs to strong coupling first as the energy scale
decreases.  Taking the Seiberg dual of the $SU(N+3M)$ group yields the
same quiver and superpotential but with $N \to N-M$, i.e. the number
of D3-branes slowly decreases with energy scale.

Before tackling the problem of what happens when $N$ becomes of order
$M$, we would like to demonstrate how the geometry of the first del
Pezzo can be recovered from the F-term relations of this gauge theory.

\subsection{The Chiral Ring}

The superpotential (\ref{WdP1}) gives rise to the ten classical F-term equations
\be
\frac{\partial W}{\partial X_i} = 0 \ ,
\ee
one for each $X_i$, where $X_i = U_L^\alpha$, $U_R^\alpha$, $V^\alpha$, $Y_j$, or $Z$.
We list these ten equations:
\begin{eqnarray}
\label{Y1}
U_L^1 V^2 &=& U_L^2 V^1 \ , \\
\label{Y2}
V^1 U_R^2 &=& V^2 U_R^1 \ , \\
\label{Y3}
U_R^1 Z U_L^2 &=& U_R^2 Z U_L^1 \ , \\
\label{Z}
U_L^1 Y_3 U_R^2 &=& U_L^2 Y_3 U_R^1 \ , \\
\label{UL}
V^\beta Y_1 &=& Y_3 U_R^\beta Z \ , \\
\label{UR}
Y_2 V^\beta &=& Z U_L^\beta Y_3 \ , \\
\label{V}
Y_1 U_L^\beta &=& U_R^\beta Y_2 \ . 
\end{eqnarray}

To see the geometry emerge from the F-term equations, 
we study the spectrum of chiral primary operators
of the schematic form
\be
{\mathcal O} = \mbox{tr} \left( X_{i_1} X_{i_2} X_{i_3} \ldots X_{i_n} \right) \ ,
\label{chpr}
\ee 
where we contract all of the color indices to form a gauge invariant
operator.  The trace structure means that these gauge invariant
operators will correspond to loops in the quiver diagram.

The reason this expression (\ref{chpr}) is schematic is that the
actual chiral primary operator will be a polynomial of such traces.
The traces generate the chiral ring.  Just as we think of harmonic
forms as representatives of classes in cohomology, we can think of a
chiral primary operator as a representative of a class of operators
related by the F-term equations.  By only giving a single trace, we
are really specifying a class of such operators related to each
other by the F-term equations.

There are six types of operators that loop just once around the quiver
\[
U_R^\alpha Y_2 Y_3 \; ; \; \; \; 
U_R^\alpha Y_2 V^\beta \; ; \; \; \;
Y_1 U_L^\alpha Y_3 \; ; \; \; \;
Y_1 U_L^\alpha V^\beta \; ; \; \; \;
\]
\be
U_R^\alpha Z U_L^\beta Y_3 \; ; \; \; \;
U_R^\alpha Z U_L^\beta V^\gamma \ .
\ee
{}From these six building blocks, we can form any operator of the form
(\ref{chpr}).  The final trace will always be over the gauge group in
the lower right hand corner of the quiver.

Now of these six building blocks, only three are independent.  Note
that from (\ref{V}), $U_R^\alpha Y_2 Y_3 = Y_1 U_L^\alpha Y_3$ and
that $U_R^\alpha Y_2 V^\beta = Y_1 U_L^\alpha V^\beta$.  Moreover,
under (\ref{UR}), $U_R^\alpha Y_2 V^\beta= U_R^\alpha Z U_L^\beta
Y_3$.  We will choose the three independent blocks
\be
U_R^\alpha Y_2 Y_3 \; ; \; \; \; 
U_R^\alpha Z U_L^\beta Y_3 \; ; \; \; \;
U_R^\alpha Z U_L^\beta V^\gamma \ .
\label{threeblock}
\ee

These blocks satisfy some additional important relations.  They are
symmetric under interchange of their $SU(2)$ indices.  From
(\ref{Y3}), we see that $U_R^\alpha Z U_L^\beta Y_3=U_R^\beta Z
U_L^\alpha Y_3$.  Moreover, from (\ref{Y3}) and (\ref{Y1}), we see
that
\be
U_R^\alpha Z U_L^\beta V^\gamma=U_R^\alpha Z U_L^\gamma V^\beta = 
U_R^\gamma Z U_L^\alpha V^\beta \ .
\ee

We now study what happens when we assemble these blocks into larger operators.
We find that
\begin{eqnarray}\label{annihilation}
U_R^\alpha Y_2 Y_3 \cdot U_R^\beta Z U_L^\gamma V^\delta 
&=& U_R^\alpha Y_2 V^\beta Y_1 U_L^\gamma V^\delta \\
&=& U_R^\alpha Z U_L^\beta Y_3 U_R^\gamma Y_2 V^\delta \nonumber \\
&=& U_R^\alpha Z U_L^\beta Y_3 \cdot U_R^\gamma Z U_L^\delta Y_3 \nonumber \ . 
\end{eqnarray}
The $\cdot$ has been added as a guide to the eye. 
Similar types of manipulation reveal also that these blocks commute with one another.
In particular
\be
U_R^\alpha Y_2 Y_3 \cdot U_R^\beta Z U_L^\gamma Y_3 = 
U_R^\beta Z U_L^\gamma Y_3 \cdot U_R^\alpha Y_2 Y_3 \ ,
\ee
and that
\be
U_R^\alpha Z U_L^\beta Y_3 \cdot  U_R^\gamma Z U_L^\delta V^\epsilon = 
U_R^\gamma Z U_L^\delta V^\epsilon  \cdot U_R^\alpha Z U_L^\beta Y_3
\ .
\ee
Finally, with a little more work, one can show that any product of 
the (\ref{threeblock}) 
is symmetric in the $SU(2)$ indices.

{}From these assembled facts, we conclude that the most general operator of the form
(\ref{chpr}) takes the schematic form
\be
\mbox{tr} \left( \prod_{i=1}^n (U_R^{\alpha_i} Z U_L^{\beta_i} Y_3)
\prod_{j=1}^m (U_R^{\gamma_j} Y_2 Y_3) \right)
\label{ch1}
\ee
or
\be
\mbox{tr} \left( \prod_{i=1}^n (U_R^{\alpha_i} Z U_L^{\beta_i} Y_3)
\prod_{j=1}^m (U_R^{\gamma_j} Z U_L^{\delta_j} V^{\epsilon_j} ) \right) \ .
\label{ch2}
\ee 
These operators are understood to be totally symmetric in the
$SU(2)$ indices.  The first operator transforms in a $2n+m+1$
dimensional representation of $SU(2)$ while the second transforms in a
$2n+3m+1$ dimensional representation.\footnote{  
\label{fn:Rcharges}
In the case $M=0$, there is a superconformal symmetry and we can assign
R-charges to these elements of the chiral ring.
The R-charge of (\ref{ch1})
is $2n + m(R_U+2R_Y)$ while the R-charge of (\ref{ch2}) 
is $2n + m(2R_U + R_Z + R_V)$.  We will recalculate these charges in 
the appendix directly from the metric.
}

At this point, it is convenient to introduce some new notation for
the building blocks (\ref{threeblock}):
\be
\begin{array}{ccc}
\left.
\begin{array}{c}
a_1 = U_R^1 Y_2 Y_3 \\
a_2 = U_R^2 Y_2 Y_3 
\end{array}
\right. ,
&
\left.
\begin{array}{c}
b_1 = U_R^1 Z U_L^1 Y_3 \\
b_2 = U_R^1 Z U_L^2 Y_3 \\
b_3 = U_R^2 Z U_L^2 Y_3 
\end{array}
\right. ,
&
\left.
\begin{array}{c}
c_1 = U_R^1 Z U_L^1 V^1 \\
c_2 = U_R^1 Z U_L^1 V^2 \\
c_3 = U_R^1 Z U_L^2 V^2\\
c_4 = U_R^2 Z U_L^2 V^2 
\end{array}
\right. .
\end{array}
\ee

{}From these nine operators, which we can treat as commuting, we can
construct any operator of the form (\ref{ch1}) or (\ref{ch2}) subject
to some relations.  In particular, we know that the resulting operator
must be totally symmetric in the $SU(2)$ indices and moreover an $a$
and $c$ type operator annihilate to form two $b$ operators
(\ref{annihilation}).  The set of twenty relations these operators
satisfy is given in Appendix A.

We can find a set of monomials with identical algebraic properties.
In particular, we let $x$, $y$, $z \in {\mathbb C}$ and
\be
\begin{array}{ccc}
\begin{array}{c}
a_1 = x^2 y \\
a_2 = x^2 z 
\end{array} ,
\begin{array}{c}
b_1 = x y^2 \\
b_2 = x y z \\
b_3 = x z^2 
\end{array} ,
\begin{array}{l}
c_1 = y^3 \\
c_2 = y^2 z \\
c_3 = y z^2 \\
c_4 = z^3 
\end{array} .
\end{array}
\label{c4}
\ee
In this way, the operators commute by construction and the number of
$x$'s labels whether the operator is of type $a$, $b$, or $c$. 
 The
number of $y$'s equals the number of $SU(2)$ indices set to one while
the number of $z$'s equals the number of $SU(2)$ indices set to two.
In any product of the $a$, $b$, and $c$, the fact that the $x$, $y$, and $z$
commute is equivalent to the fact that the operator is totally
symmetric in the $SU(2)$ indices.  Only slightly more difficult to see
is (\ref{annihilation}).

Because the $a_i$, $b_j$ and $c_k$ commute, one can diagonalize all of
these matrix valued operators simultaneously. Eigenvalue by
eigenvalue, these operators describe a three dimensional variety,
parametrized above by $x,y,z$. This variety is the Calabi-Yau geometry
we are interested in. The description of a general point in the moduli
space is a set of $N$ unordered points of the Calabi-Yau geometry. We
call this moduli space the mesonic branch of the theory. That we have
found a Calabi-Yau in this way is in line with the expectations of
reverse geometric engineering \cite{Berensteinrev}.

Now, treating $x$, $y$, and $z$ as homogenous coordinates on ${\mathbb
  P}^2$, this set of monomials can be reinterpreted as a set of
linearly independent polynomials on ${\mathbb P}^2$ which vanish at
the point $(1,0,0) \sim (\lambda,0,0)$ for $\lambda \in {\mathbb
  C}^*$.  This set of monomials then provides a map from ${\mathbb
  P}^2 \to {\mathbb P}^8$ with the point $(1,0,0)$ missing because the
origin is not contained in ${\mathbb P}^8$.  The smallest surface
containing the image of this map is well known to be the first del
Pezzo, or ${\mathbb P}^2$ blown up at a point.

\subsection{The First del Pezzo}

We can be more precise in our description of this first del Pezzo.  It
is easy to embed ${\mathbb P}^2$ blown up at a point inside ${\mathbb
  P}^2 \times {\mathbb P}^1$.  Let $x$, $y$, and $z$ be homogenous
coordinates on the ${\mathbb P}^2$, as above, and let $w_1$ and $w_2$
be coordinates on the ${\mathbb P}^1$.  The first del Pezzo is
described as the hypersurface $dP_1 \subset {\mathbb P}^2 \times
{\mathbb P}^1$ satisfying $y w_2 = z w_1$.

We now construct a map from ${\mathbb P}^2 \times {\mathbb P}^1$ to
${\mathbb P}^{11}$ by considering all monomials linear in the $w_i$
and quadratic in the $x$, $y$, and $z$.  (Algebraic geometers would
call this map a composition of the Segre and Veronese maps.)
\begin{eqnarray*}
{\mathbb P}^2 \times {\mathbb P}^1 &\longrightarrow& {\mathbb P}^{11} \\
(x,y,z; w_1, w_2) &\longmapsto& (x^2 w_1, xy w_1, xz w_1, y^2 w_1, yz w_1, z^2 w_1, \\
&&  \;  x^2 w_2, xy w_2, xz w_2, y^2 w_2, yz w_2, z^2 w_2)
\end{eqnarray*}
This map restricts to an injective map from $dP_1$ to ${\mathbb P}^8$
by requiring $y w_2 = z w_1$ -- of the twelve monomials above, only
nine are linearly independent under this relation.

Now for a point on $dP_1$ where $y$ and $z$ do not both vanish, the
constraint $y w_2 = z w_1$ fixes a point on the ${\mathbb P}^1$ which
we can take to be $w_2 = z$ and $w_1=y$.  In this case, the twelve
monomials in ${\mathbb P}^{11} $ reduce precisely to the nine
monomials given as (\ref{c4}).  Thus this $dP_1 \subset {\mathbb P}^8$
certainly contains the image of the map described above from ${\mathbb
  P}^2 \to {\mathbb P}^8$.

Using the twelve monomials that depend on the $w_i$, we can also
describe the extra blown up ${\mathbb P}^1$ in $dP_1$.  Consider the
case $y=0$ and $z=0$.  Now, the $w_i$ are unconstrained, and the only
two nonvanishing monomials in the ${\mathbb P}^{11}$ are $x^2 w_1$ and
$x^2 w_2$, which parametrize the blown up ${\mathbb P}^1$.

Naively, this description of the first del Pezzo seems overly
complicated.  For example, we may just as easily consider the simpler
Segre map to ${\mathbb P}^5$:
\begin{eqnarray*}
{\mathbb P}^2 \times {\mathbb P}^1 &\longrightarrow& {\mathbb P}^{5} \\
(x,y,z; w_1, w_2) &\longmapsto& (x w_1, y w_1, z w_1, x  w_2, y w_2, z w_2)
\end{eqnarray*}
Subject to the relation $yw_2 = zw_1$ we have a simpler embedding of
the first del Pezzo inside ${\mathbb P}^4$.  However, we are not just
describing the geometry of the Del-Pezzo surface, but a particular
algebraic cone over it, which results from blowing down the zero
section of the total space of a line bundle over $dP_1$ which makes
the space a Calabi-Yau geometry.

\section{D5-branes and Superconformal Fixed Points}

We will now give a plausible argument that the field theory on a stack
of D3-branes and D5-branes probing a $dP_1$ singularity does not have
a superconformal fixed point in the infrared. This absence helps rule
out strong coupling behaviors in the IR field theory which are not
gaugino condensation.

A necessary condition for the presence of a superconformal fixed point
in a gauge theory is the vanishing of the NSVZ beta functions and the
requirement that the superpotential be marginal.  We will assume that
there is an $SU(2)$ global symmetry.  As a result, the anomalous
dimensions of the doublet fields will be equal, $\gamma_{V^1} =
\gamma_{V^2}$ and similarly for the $U_L$ and $U_R$.  Moreover,
because of this $SU(2)$, there will only be three independent
superpotential couplings.

We find it convenient to work with R-charges rather than with the
anomalous dimensions $\gamma$.  Recall that at a superconformal fixed
point, the supersymmetry algebra implies that $3R = 2+ \gamma$.

For the NSVZ beta functions to vanish, the following expressions must vanish:
\begin{eqnarray*}
\beta_0 &=& 2N + (N+3M) (R_Z - 1) + (N+M) (R_{Y_1} - 1) + 2(N+2M) (R_{U_L} - 1) \ , \\
\beta_1 &=& 2(N+M) + (N+2M) (R_{Y_3}-1) + 2(N+2M)(R_V -1) + \\
&& N (R_{Y_1} -1)+2(N+3M)(R_{U_R}-1) \ , \\
\beta_2 &=& 2(N+2M) + (N+M)(R_{Y_3}-1) + 2(N+M)(R_V-1) +\\
&& (N+3M)(R_{Y_2}-1) + 2N (R_{U_L}-1) \ , \\
\beta_3 &=& 2(N+3M) + N (R_Z-1) +(N+2M)(R_{Y_2}-1) + 2(N+M) (R_{U_R}-1) \ .
\end{eqnarray*}
Moreover, the superpotential couplings must be marginal, which implies that
each of the $b_i$ have R-charge two:
\begin{eqnarray*}
2 &=& R_{U_L} + R_{Y_1} + R_V \, \\
2 &=& R_{U_R} + R_{Y_2} + R_V \ , \\
2 &=& R_Z + R_{U_R}  + R_{Y_3} + R_{U_L} \ .
\end{eqnarray*}
This set of seven equations is not linearly independent.  In the case $M=0$,
there is a two parameter family of solutions.  In the case $M \neq 0$, there
is just a one parameter family.  

However, the existence of a family of solutions to these beta function
constraints is not sufficient to guarantee the existence of a
superconformal fixed point.  From Intriligator and Wecht \cite{IW}, we
know that at a superconformal fixed point, the conformal anomaly $a
\sim {\mbox Tr} R^3$ should be at a local maximum.  Here, ${\mbox Tr}
R^3$ is a trace over the R-charges of all of the fermions in the gauge
theory.  For the $M=0$ case, it is indeed possible to maximize $a$
over the two parameter family of solutions, and one can match the
results to geometric expectations \cite{Martelli, Bertolini,
  Benvenuti}.  However, in the case $M \neq 0$, $a$ cannot be
maximized.  If we let $R_V$ parametrize the family of solutions to the
beta function constraints, it is straightforward to see that
\be
\tr R^3 = 24( -19 M^2 - 27 MN - 9N^2 + (M+N)(2M+N) R_V)
\ee
is a linear function in $R$.

In fact, one can see the pathology in the $M\neq 0$ case even before
computing $a$.  The R-charges of the $a_i$, $b_j$ and $c_k$ operators
of the previous section do not depend on $R_V$.  One finds that
$R_{c_k} = 6$ while $R_{a_i} = -2$.  In other words, a finite fraction
of operators in the chiral ring -- those with enough of the $a_i$ as
building blocks -- will have a conformal dimension below the unitarity
bound.

Such a pheonomenon has been encountered before studying simpler SQCD
like gauge theories \cite{Kutasov, Seibergduality} where it was argued
that accidental U(1) symmetries emerge in the infrared which decouple
these troublesome operators from the theory once their dimension
reaches the unitarity bound.  However, in these simpler cases, there
was a finite number of such operators and there was also a well
defined ultraviolet where the theory became asymptotically free.  We
have an infinite number of troublesome operators, and truncating them
from the theory will radically change the geometry.

A simpler and we believe more probable resolution to this pair of
problems -- an inability to maximize $a$ and an infinite set of
operators below the unitarity bound -- is that the corresponding
superconformal fixed point does not exist.

At the beginning, we assumed an SU(2) symmetry.  If we allow this
SU(2) to be broken, we can maximize $a$ but there will still be an
infinite number of operators with negative conformal dimension.

\section{The Quantum Moduli Space}
\label{quantummoduli}

We begin by analyzing the theory with no D3-branes and $N$ D5-branes.
Such a gauge theory can be thought of as the last step in the Seiberg
duality cascade, starting in the UV with a multiple $nN$ of D3-branes.
Starting with the pure D5-brane theory precludes the possibility of a
baryonic branch, a possibility we will return to later.

The pure D5-brane gauge theory has gauge group $U(3N)\times U(2N)
\times U(N)$ and classical superpotential
\be
W = U_R^2 Y_2 V^1 - U_R^1 Y_2 V^2 \ .
\ee
We will keep the $SU(2)$ global symmetry, and thus single loops
involving the $Y_3$ field cannot appear in the superpotential.  
Most aspects of our analysis are standard in the study of 
${\mathcal N}=1$ supersymmetric gauge theories.  See 
\cite{reviews} for a selection of relevant reviews.

First, we argue that confinement and gluino condensation occur at low
energies in our gauge theory.  Because these gauge groups are $U(N)$,
they will have three decoupled $U(1)$ subgroups, some of which allow
the addition of Fayet-Iliopoulos (FI) terms.\footnote{ An FI term for
  the $U(3N)$ group is forbidden.  When we write out the full matrix
  structure of the D-term equations, the D-term equation for the
  $U(3N)$ gauge group is of the form $\zeta \delta^i_j = (Y_2)^i_a
  (Y_2^\dagger)^a_j - (U_R^{\alpha\dagger})^i_m (U_R^\alpha)^m_j$ with
  the $Y_2$ field contracted on its $U(2N)$ indices and the $U_R$
  fields contracted on their $U(N)$ indices.  For positive (negative)
  $\zeta$ we may consider giving $Y_2$ ($U_R$) an expectation value
  independent of the other field.  But by applying gauge rotations to
  this field it is always possible to eliminate components of the
  field such that the contracted form is diagonal and nonvanishing
  only on $2M$ of its $3M$ diagonal components, inconsistent with the
  D-term equations.}
Suppose we add an FI term $\kappa$ for the $U(1) \subset U(2N)$.  For
the purposes of this argument we take $\kappa$ to be large compared to
other scales in the problem and temporarily ignore the superpotential.
The relevant piece of the classical K\"ahler potential for the gauge
theory is
\be
(\kappa + V^1 (V^1)^\dagger + V^2 (V^2)^\dagger + Y_3 Y_3^\dagger - 
Y_2 Y_2^\dagger )\ .
\ee
For negative $\kappa$ we may give expectation values to $V^1$ and $V^2$.  

Now, consider the D-term equation associated with the $U(N)$ gauge group:
\be
(\zeta + U_R^1(U_R^1)^\dagger + U_R^2(U_R^2)^\dagger -  V^1 (V^1)^\dagger - V^2 (V^2)^\dagger -Y_3 Y_3^\dagger) \ .
\ee
We choose the FI term $\zeta=-2\kappa$ and set the expectation values
of $U_R^\alpha$ and $Y_2$, $Y_3$ equal to zero.

Turning now to the superpotential, we see that the large expectation
values for the $V^\alpha$ give large masses to the $U_R^\alpha$ and
$Y_2$ superfields, which we may then integrate out.  But after
integrating out these fields, we see that the $U(3N)$ group decouples
from the other gauge groups, and we know that pure $U(3N)$ gauge
theory with no flavors generically confines and that the gauginos
condense.

Although we have carried out the analysis for large
$\kappa$, we can now tune $\kappa$ away to zero.  Since confinement
and chiral symmetry breaking
are low energy phenomena controlled by F-terms, we expect their
presence or absence to be independent of $\kappa$.
Therefore we expect confinement and chiral symmetry
breaking for our triangle quiver gauge theory.

This analysis applies to the general point in the mesonic branch, with
or without extra probe D3-branes in the bulk.  If we had additional
branes in the bulk, confinement of the fractional brane stuck at the
singularity should still happen. There is a lot of evidence that
branes in the bulk do not modify the essential dynamics of the
fractional branes \cite{Berensteinmatrix}, although the fractional
branes do modify the dynamics of the branes in the bulk by deforming
their moduli space. In essence, the analysis of the fractional branes
can be done without ever mentioning the branes in the bulk, but the
result will be correct even if branes in the bulk are present.  If
gaugino condensation happens, we should expect to see a deformed
moduli space for the branes in the bulk: the quantum moduli space of
the gauge theory.

An important point is that our
$SU(3N)\times SU(2N) \times SU(N)$ theory has no flat mesonic
directions.  The F-term equations for the classical superpotential
imply that $U_R^\alpha Y_2 = 0$.  Thus, we cannot form any nonzero
gauge invariant trace operators.  There are some anti-symmetrized
products of fields, the dibaryon operators, for example
\be
{\mathcal B} = \epsilon_{\alpha_1 \cdots \alpha_{2N}} \epsilon^{\beta_1 \cdots \beta_N}
\epsilon^{\gamma_1 \cdots \gamma_N}
(Y_3)^{\alpha_1}_{\beta_1} \cdots (Y_3)^{\alpha_{N}}_{\beta_N}
(V^1)^{\alpha_{N+1}}_{\gamma_1} \cdots (V^1)^{\alpha_{2N}}_{\gamma_N} \ .
\ee
By turning on FI terms 
for the $U(2N)$ and $U(N)$ gauge groups, we can fix the expectation value of ${\mathcal B}$.

\subsection{SUSY Breaking for Strongly Coupled $SU(3N)$}
\label{simpleSUSYb}

Ignoring the $U(1)$ subgroups for the moment, note that the $SU(3N)$ group is
the only one of the three gauge groups to have $N_f < N_c$, so it is
natural to suppose that it runs to strong coupling first.  In this
case we may construct the effective superpotential at low energies by
standard arguments.  Ignoring strong coupling effects from the other
gauge groups, the quantum modified superpotential is
\cite{Seibergduality} 
\be 
W = M^2 V^1 - M^1 V^2 + N \left(\frac{\Lambda_3^{7N}}{\det M} \right)^{1/N}
\label{quantumW}
\ee 
where we have defined the mesonic operators $M^\alpha = U_R^\alpha
Y_2$ and a square matrix $M = (M^1 , M^2)$.  At strong coupling we
treat the mesons $M$ as if they were fundamental fields.

The equation of motion for the $V^\alpha$ sets $M^i = 0$.  However,
the equation of motion for the $M^i$ tells us that
\be
0 = \epsilon_{ij} V^j_{ab}  - \frac{\Lambda_3^7 \minor(M^i_{ab})}{(\det M)^{(N+1)/N}} \ .
\ee
This equation is not consistent with setting the $M^i=0$ and we conclude
that supersymmetry is broken.

Because the $SU(3N)$ theory has the largest number of colors and
smaller number of flavors, we generically expect $SU(3N)$ to run to
strong coupling first.  It is possible that the other two groups run
to strong coupling as well, which might invalidate the analysis done
above. In the next subsection an exact argument using the generalized
Konishi anomaly demonstrates that the $SU(N)$, $SU(2N)$, and $SU(3N)$
groups cannot be strongly coupled at the same time.

\subsection{Konishi Anomaly Equations}

We have argued that supersymmetry is broken when $\Lambda_3$ is the
dominant scale in the problem.  As the couplings become large,
anomalous dimensions can also become large and in the absence of a
superconformal fixed point it is very difficult to control the RG
flow.  At generic points in the space of coupling constants, the
approximations made in section \ref{simpleSUSYb} may be invalid and it
is not clear whether supersymmetry will be broken.

To make a stronger argument we need a better method to determine the
possible quantum modifications to $W$.  A technique that we will find
very powerful is based on the Konishi anomaly \cite{Konishi}; see also the review of Amati et al \cite{reviews} and the more recent developments of \cite{CDSW}. Some of this analysis can be rephrased in
terms of the language of the Dijkgraaf-Vafa matrix model approach
\cite{DV}, which is a useful way of encoding the same set of quantum
modified equations of motion.  These matrix models and Konishi
anomaly calculations can encode the (quantum) deformations of complex
structure of the Calabi-Yau space \cite{Berensteinmatrix}.

The generalized Konishi anomaly equations are derived by considering
infinitesimal variations of the fields, which leave the path integral
invariant, and obtaining the associated Ward identities, restricted to
the chiral ring.  For a variation of the bifundamental $X \to X +
\delta X$ where we take $\delta X = f(X, W_\alpha W^\alpha)$ and
$W_\alpha$ is the supersymmetric gauge field strength, one finds
\be
\left\langle -\frac{1}{32\pi^2}
\sum_{i,j} \left\{ W_\alpha, \left[W^\alpha, 
\frac{\partial f} {\partial {X_i^{\, j}} }\right] \right\}_i^{\, j} \right \rangle = 
\left \langle \tr  (f(X,W_\alpha W^\alpha) \partial_X W) \right \rangle \ .
\ee
The particular gauge group under which $W^\alpha$ transforms is fixed
by gauge invariance, which in practice corresponds to whether $W^\alpha$ appears on
the left or right side of $f$.  Inside a trace, we have the convenient result that $\tr(\phi [W_\alpha W^\alpha, X]) = 0$, where $\phi$ is some arbitrary combination of bifundamentals.  It is also useful to recall that $\{W_\alpha, W_\beta\}=0$ in the chiral ring.

Let us consider the Konishi equations for $\delta X =X$.  On varying the fields appearing in the superpotential, we find that
\begin{align}
 \expect{\tr(U_R^1Y_2V^2)}&= N S_3 + 3N S_1 = 3N S_2 + 2N S_3\ , 
\nonumber \\
 \expect{\tr(U_R^2Y_2V^1)}&= - N S_3 - 3N S_1 = -3N S_2 - 2N S_3\ , 
\label{linear} \\
\expect{\tr(U_R^2Y_2V^1)}-\expect{\tr(U_R^1Y_2V^2)} 
&= -2N S_1 - N S_2. \ \nonumber
\end{align}
where we have defined $S_k = - \tr_k W_\alpha W^\alpha / 32 \pi^2$, and
the trace is over the gauge group with rank $kN$.  There is an additional anomaly equation from varying $Y_3$, which does not appear in the superpotential:
\be
2N S_1 + N S_2 = 0.
\label{linearY3}
\ee
These equations taken collectively give relations amongst the $S_k$, and it is easy to verify that they have no solution other than the trivial one, with all the $S_k=0$.  The anomaly equations require that the infrared gauge theory does not undergo gaugino condensation.  However, this is inconsistent with the D-term arguments given earlier, which implied that some of the $S_k$ have to be non-zero.  Our anomaly arguments assumed supersymmetry, so we conclude that supersymmetry must be broken 
(or that the vacuum of the theory is in some other way ill-defined.)

The inability to satisfy these constraints is reminiscent of the geometric obstruction to deformation discovered by Altmann \cite{Altmann} (see Appendix A for a brief review).  The cone over $dP_1$ can be expressed in terms of twenty (dependent) equations in nine variables.  It is possible to deform these equations by a parameter $\epsilon$, but one finds that mutual consistency of the deformed equations requires that $\epsilon^2$ vanishes.
If $\epsilon$ were not obstructed, we would identify
this deformation of the geometry with the quantum deformed moduli
space of the mesonic branch: gaugino condensation leads to a quantum
modified chiral ring, and we have already described in the previous
section how the chiral ring and the Calabi-Yau geometry are related.
Our analysis suggests a strong relation between the geometry of
the Calabi-Yau space and spontaneous SUSY breaking. We correlate the
geometric deformation obstruction to inconsistencies of the quantum
modified chiral ring. Thus the name SUSY-BOG: supersymmetry breaking
by obstructed geometries.

For completeness let us consider another set of anomaly equations, which we obtain from the variations $\delta X = X \, W_\alpha W^\alpha$.  
The anomaly equations for the $U_R^\alpha$ and the $V^\beta$ give
\begin{align}
-\frac{1}{32\pi^2} \expect{\tr(W_\alpha W^\alpha \, U_R^1Y_2V^2)}&=-S_1S_3=- S_1 S_2 \ , 
\nonumber \\
-\frac{1}{32\pi^2} \expect{\tr(W_\alpha W^\alpha \, U_R^2Y_2V^1)}&= S_1 S_3= S_1 S_2 \ ,
\label{quad1}
\end{align}
while for the $Y_2$, we find
\begin{equation}
-\frac{1}{32\pi^2} \left(\expect{ \tr(W_\alpha W^\alpha U_R^2Y_2V^1)}-\expect{\tr(W_\alpha W^\alpha 
U_R^1Y_2V^2)} 
\right)=S_2 S_3 \ ,
\label{quad2}
\end{equation}
A similar equation for $Y_3$ yields
\be
S_1 S_2 = 0 \ .
\label{quadraticrel}
\ee
because $Y_3$ does not appear in the superpotential. 

Taking these quadratic relations for the $S_k$ on their own, we would conclude that two of the $S_k$ are zero.
Furthermore, if only one of the gauge groups has a nonzero $S_k$, we
can assume for that particular gauge group $N_f < N_c$.  The only
reasonable possibility is $S_3 \neq 0$ -- but we have already shown
that if the $SU(3N)$ group flows to strong coupling first, then
supersymmetry should be broken.

\subsection{The Baryonic Branch}

All our analysis so far has concerned the pure D5-brane theory, but
as argued before, this analysis extends to the case where there are branes
in the bulk: what happens at the singularity is independent of the branes in 
the bulk and supersymmetry is broken on the mesonic branch of the general theory.

However, there are other baryonic branches in moduli space which we
have not analyzed yet.  For our quiver theory, we can halt our Seiberg
duality cascade one step up from the bottom, at the $SU(4N)\times
SU(3N) \times SU(2N) \times SU(N)$ theory.  If the group with the
largest number of colors $SU(4N)$ runs to strong coupling first, we
have effectively $N_f = N_c$ SQCD with a tree level superpotential.
As was noticed in \cite{HEK}, such a theory will have a quantum
modified superpotential.  In addition to the mesonic branch, there is
a branch of moduli space where baryonic operators get expectation
values.

We will argue that supersymmetry is broken even if we give these
baryonic operators expectation values.  To
to try to gain a deeper understanding of the various possibilities, it is
useful to review what happens in the case of the conifold.  
One step up from the bottom of the duality cascade on the conifold,
we have an $SU(2N) \times SU(N)$ theory with bifundamentals
$A_i$ and $B_i$, $i=1,2$ and a quartic superpotential
\be
W_{tree} = \epsilon_{ij} \epsilon_{kl} A_i B_k A_j B_l
\ee
At strong coupling for the $SU(2N)$ group, 
we form the mesonic operators $M_{ij} = A_i B_j$, and
the superpotential is modified by quantum effects \cite{Seibergqm}
\be
W_{eff} = W_{tree} + X ( \det M + {\mathcal B} \tilde {\mathcal B} - \Lambda_{2N} ) \ .
\label{Weffnfnc}
\ee
On the baryonic branch, $\langle {\mathcal B} \tilde {\mathcal B}
\rangle = \Lambda_{2N}$ while the mesonic expectation values and the
Lagrange multiplier $X$ vanish.  One may obtain a low-energy effective
theory on this branch by integrating out the mesons, leaving us with a
pure $SU(N)$ gauge theory supplemented by the baryonic operators
${\mathcal B}, \tilde{\mathcal B}$, which lie on a flat direction.

The Klebanov-Strassler supergravity solution corresponds to the
symmetric point ${\mathcal B} = \tilde {\mathcal B}$ on the baryonic
branch \cite{GHK, Ofer}.  A one real parameter family of
supersymmetric supergravity solutions corresponding to changing
$|{\mathcal B}|$ was worked out by \cite{Butti}.  The remarkable fact
about this one parameter family of solutions is that it does not
involve a Calabi-Yau manifold.  Instead, it requires only a manifold
with $SU(3)$ structure.

The Altmann theorem mentioned in the introduction states only that the
Calabi-Yau cone over $dP_1$ has no complex structure deformations.  If
we relax the Calabi-Yau condition to an $SU(3)$ structure condition,
there may still be supersymmetric supergravity solutions, as there are
for more general points on the baryonic branch of the conifold theory.

We now exclude the possibility of an $SU(3)$ structure solution.
Returning to the four-node quiver theory with gauge group
$SU(4N)\times SU(3N)\times SU(2N) \times SU(N)$, with the $SU(4N)$
group at strong coupling, we argue that although this theory has a
baryonic branch the Konishi anomaly arguments still imply that
supersymmetry is broken.  The first step is to replace the elementary
fields charged under the $SU(4N)$ gauge group by mesons and baryons
which are $SU(4N)$ singlets.  Specifically, we construct mesons
$M_Y^\alpha = U_R^\alpha Y_2$ and $M_Z^\alpha = U_R^\alpha Z$, which
have their $SU(4N)$ indices contracted, and baryons ${\mathcal B}=
[U_R^1\cdots U_R^1 U_R^2 \cdots U_R^2]$ and $\tilde{\mathcal B} = [Y_2
  \cdots Y_2 Z \cdots Z]$, whose $SU(4N)$ indices are fully
antisymmetrized.  This theory develops a quantum-modified
superpotential of the form (\ref{Weffnfnc}) and has a baryonic branch,
along which the baryons acquire expectation values.  On this branch
one also finds that the Lagrange multiplier $X$ and $\det M$ vanish.

What happens to the remaining mesons?  The mesons $M_Z$ appear only in
a cubic term in the tree-level superpotential, so they remain as light
fields in the theory.  On the other hand, the mesons $M_Y$ have a mass
term from the tree-level superpotential of the form $V^\alpha
M_Y^\beta \epsilon_{\alpha\beta}$ and we may integrate these fields
out.  The F-terms for the mesons $M_Y$ force $V^\alpha=0$, while the
F-terms from varying $V^\alpha$ require $M_Y^\alpha = Y_1 U_L^\alpha$.
By applying gauge rotations to the $Y_1$ and $U_L$ fields one can see
that the determinant of the meson matrix vanishes, and that the F-term
equations are therefore satisfied consistently on this branch.

After integrating out these massive fields, and relabeling the light
fields, one finds that the resulting low-energy effective theory is
precisely that of the three-node quiver, supplemented by the baryonic
operators.  However, the baryons do not communicate with the
$SU(3N)\times SU(2N) \times SU(N)$ degrees of freedom.  This baryonic
flat direction cannot interfere with the arguments for supersymmetry
breaking given in section \ref{quantummoduli}.

\subsubsection*{Probe D3-branes}

The preceding argument about supersymmetry breaking along the baryonic
branch was not as strong as the Konishi anomaly arguments about the
mesonic branch.  For example, we were not able to analyze the
possibility that both the $SU(3N)$ and $SU(4N)$ groups ran to strong
coupling at once.  Hedging our bets, we will assume for the moment the
existence of a supersymmetry preserving baryonic branch for the $dP_1$
theory and a corresponding $SU(3)$ structure supergravity solution.

We may ask what happens to probe D3-branes in this putative
supersymmetric vacuum.  After all, for generic initial numbers of D3-
and D5-branes in the UV, the duality cascade will generically leave
some left over D3-branes in the bulk.  In the Klebanov-Strassler
solution, these extra D3-branes were supersymmetric, but what happens
at an arbitrary point on the baryonic branch?

Away from the KS point, these probe D3-branes in the conifold should
break supersymmetry.  The authors of \cite{Butti} claim that this
baryonic branch supergravity solution for the conifold interpolates
between the KS and the Maldacena-Nunez (MN) \cite{MN} supergravity
solutions.  Moreover, for MN, D3-brane probes break supersymmetry.

Let us describe more explicitly these supersymmetry conditions for
D3-brane probes.  Let $\epsilon$ be a Killing spinor in the ten
dimensional geometry.  The presence of ${\mathcal N}=1$ supersymmetry
in four dimensions guarantees the existence of at least a four
dimensional representation of such spinors.  For a probe D3-brane to
be supersymmetric, $\epsilon$ must be an eigenspinor of the
corresponding $\kappa$-symmetry matrix for the probe: $\Gamma_\kappa
\epsilon = \epsilon$.  For a D3-brane aligned in the gauge theory
directions, $x$, $y$, $z$, and $t$, the $\kappa$-symmetry matrix is
just the four dimensional gamma matrix
\be
\Gamma_\kappa = i \Gamma_t \Gamma_x \Gamma_y \Gamma_z \ .
\ee
In other words, $\epsilon$ must have positive chirality in the four
gauge theory directions.

The Killing spinors in the \cite{Grana} $SU(3)$ structure solutions
take the general form
\be
\epsilon = \alpha \zeta_+ \otimes \eta_+ + \beta \zeta_- \otimes \eta_- \ ,
\ee
where $\zeta_\pm$ are the chiral and anti-chiral four dimensional spinors, and $\eta_\pm$ are chiral and anti-chiral six dimensional spinors.  The $\alpha$ and $\beta$ are complex functions of coordinates on the six-dimensional transverse geometry. In the KS solution, $\beta = 0$, and so $\Gamma_\kappa \epsilon= \epsilon$, while for the MN solution, $\beta = -i \alpha^*$, and the D3-brane breaks supersymmetry.  Indeed, for any nonzero value of $\beta$, the D3-brane will break supersymmetry.

Using the table of possible $SU(3)$ structure solutions in \cite{Grana}, we can make a stronger, more general statement.  If the D3-brane preserves supersymmetry, then $\beta=0$.  If $\beta=0$, the six dimensional transverse geometry must be Calabi-Yau.  However, Altmann claims there is no Calabi-Yau deformation of the cone over $dP_1$.  Therefore, if there is an $SU(3)$ structure solution corresponding to the baryonic branch of this $dP_1$ field theory, probe D3-branes will break supersymmetry.

\section{Outlook}

We have shown, using a combination of traditional field theory and
supergravity arguments, that generically supersymmetry is
spontaneously broken at low energies in the duality cascade
constructed from the cone over $dP_1$.  Traditional field theory
convinced us that the mesonic branch must break supersymmetry.
Although there is a small possibility that the baryonic branch remains
supersymmetric, for generic initial numbers of D3- and D5-branes,
there will be extra D3-branes at the bottom of the cascade which we
argued should also lead to supersymmetry breaking.

This mechanism for supersymmetry breaking by obstructed geometries, or
SUSY-BOG, should work more generally.  We start with some Calabi-Yau
cone to which we add D3- and D5-branes.  Generically, a duality
cascade will result, where the number of D3-branes is gradually
reduced as we move toward the tip of the cone.  When we reach the bottom of the cascade, 
we expect to find a confining theory and presumably confinement will happen via gaugino 
condensation. In such a situation, we expect a deformed chiral ring and hence a deformed 
complex structure of the geometry.\footnote{
See \cite{urangapapers} for more examples of unobstructed deformations where the 
endpoint of the cascade is either a confining theory or a superconformal fixed point.}
The unifying feature
of the SUSY-BOG mechanism is an obstruction in the complex deformation
space of the Calabi-Yau cone.  Our conjecture is that in all of these 
situations supersymmetry will be spontaneously broken. The cone over $dP_1$ is
 not alone in
having such an obstruction.  For example, we expect to be able to set
up a duality cascade using the $Y^{p,q}$ space \cite{HEK}, but there
is such an obstruction for all the $Y^{p,q}$, $p>q>0$.

For the higher del Pezzos $dP_n$, $1<n<9$, there will also generically
be a problem.  We expect to be able to set up a duality cascade.  For
the general $dP_n$ there are $n$ vanishing two-cycles and thus $n$
types of D5-branes.  However, in general there are fewer than $n$
complex structure deformations \cite{CFIKV}.  For example, for $dP_2$,
there is only one such deformation.  Thus, unless the initial numbers
of D5-branes are chosen carefully, we expect SUSY-BOG to occur at low
energies, at the end of the duality cascade.

While we were finishing this paper, \cite{Burrington} appeared.  The
authors claim to find a first order complex structure deformation of
the cone over $dP_1$ and construct from this first order deformation
the analog of the KS solution \cite{KS} for $dP_1$.  This construction
is completely compatible with our story.  We anticipate, from
\cite{Altmann}, that they will not be able to continue their
deformation to higher order in $\epsilon$.  It would be very interesting, but 
presumably quite difficult, to construct a non-supersymmetric gravity 
solution for the deformed cone.

{\bf Note Added} 
After this work was completed, we became aware of the closely related
papers \cite{scoop, scoop2} which reach similar conclusions.

\section*{Acknowledgments}

We would like to thank Marcus Berg, Aaron Bergman, Henriette Elvang,
Michael Haack, Igor Klebanov, Matt Strassler, and Brian Wecht for
useful discussions and comments.  P.O. thanks the University of
Washington Particle Theory Group for hospitality while this work was
being completed. The work of D.~B. and S.~P is supported in part by
DOE grant DE-FG02-91ER40618. The research of C.~P.~H. is supported in
part by the NSF under Grant No.~PHY99-07949.  The work of P.~O. is
supported in part by the DOE under grant DOE91-ER-40618 and by the NSF
under grant PHY00-98395.

\appendix

\section{Relations for the first del Pezzo}

The twenty relations among the $a_i$, $b_j$, and $c_k$ for the first del Pezzo are
\be
\begin{array}{cccc}
b_2^2 = a_1 c_3 &
b_2^2 = b_1 b_3 &
b_2^2 = a_2 c_2 &
c_2^2 = c_1 c_3 \\
c_3^2 = c_2 c_4 &
b_1^2 = a_1 c_1 &
b_3^2 = a_2 c_4 &
b_1 c_2 = b_2 c_1 \\
b_1 c_3 = b_2 c_2 &
b_1 c_4 = b_2 c_3 &
b_2 c_3 = b_3 c_2 &
b_2 c_4 = b_3 c_3 \\
a_1 b_2 = a_2 b_1 &
a_1 b_3 = a_2 b_2 &
b_2 c_2 = b_3 c_1 &
a_1 c_2 = b_1 b_2 \\
a_1 c_4 = b_2 b_3 &
c_1 c_4 = c_2 c_3 &
a_2 c_1 = b_1 b_2 &
a_2 c_3 = b_2 b_3 
\end{array}
\ee

Because the cone over $dP_1$ is a toric variety, there is an easy way to summarize these
relations.
Given a set of vectors $f_i$ which generate the lattice inside the dual toric cone
$\sigma^\vee$, the set of relations among the coordinate ring is easily summarized
as the set of integer valued vectors $\vec m_I$ such that
\be
\sum_i f_i m^i_I = 0 \ .
\ee
For the first del Pezzo, we can read off twenty such relations from 
Figure \ref{sigmavdp1}:

\begin{figure}
\begin{center}
\includegraphics[width=2in]{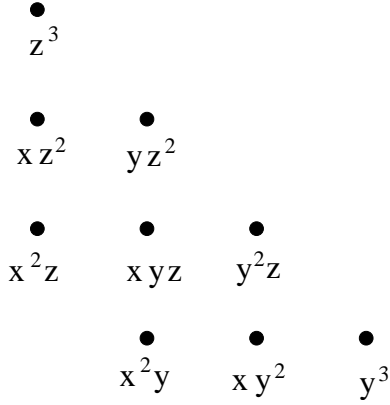}
\end{center}
\vfil
\caption{The lattice points which generate the dual cone
for ${\mathbb P}^2$ blown up at a point.
}
\label{sigmavdp1}
\end{figure}

Altmann claims \cite{Altmann} that the one dimensional space of complex structure
deformations is obstructed at second order.  One can modify the above twenty
equations by adding $\epsilon$:
\be
\begin{array}{llll}
b_2 (b_2 - 2 \epsilon) = a_1 c_3 &
b_2 (b_2 - 3 \epsilon) = b_1 b_3 &
b_2 (b_2 - 4 \epsilon) = a_2 c_2 &
c_2^2 = c_1 c_3 \\
c_3^2 = c_2 c_4 &
b_1^2 = a_1 c_1 &
b_3^2 = a_2 c_4 &
b_1 c_2 = (b_2 -\epsilon) c_1 \\
b_1 c_3 = (b_2-\epsilon) c_2 &
b_1 c_4 = (b_2-\epsilon) c_3 &
(b_2-2\epsilon) c_3 = b_3 c_2 &
(b_2-2\epsilon) c_4 = b_3 c_3 \\
a_1 (b_2 - 3\epsilon) = a_2 b_1 &
a_1 b_3 = a_2 b_2  &
(b_2-2 \epsilon) c_2 = b_3 c_1 &
a_1 c_2 = b_1 (b_2-\epsilon) \\
a_1 c_4 = b_2 b_3 &
c_1 c_4 = c_2 c_3 &
a_2 c_1 = b_1 (b_2- 3 \epsilon ) &
a_2 c_3 = (b_2 -2\epsilon) b_3 
\end{array}
\ee
However, for consistency, $\epsilon^2=0$.

\section{Gauge Theories and Chiral Rings for the $Y^{p,q}$}
\label{sec:dual}

In studying the chiral ring for $dP_1$, we found it was easy to generalize to the more
complicated quivers dual to the $Y^{p,q}$ spaces.  We include these results in this
appendix.

\begin{figure}
\begin{center}
\includegraphics[width=2in]{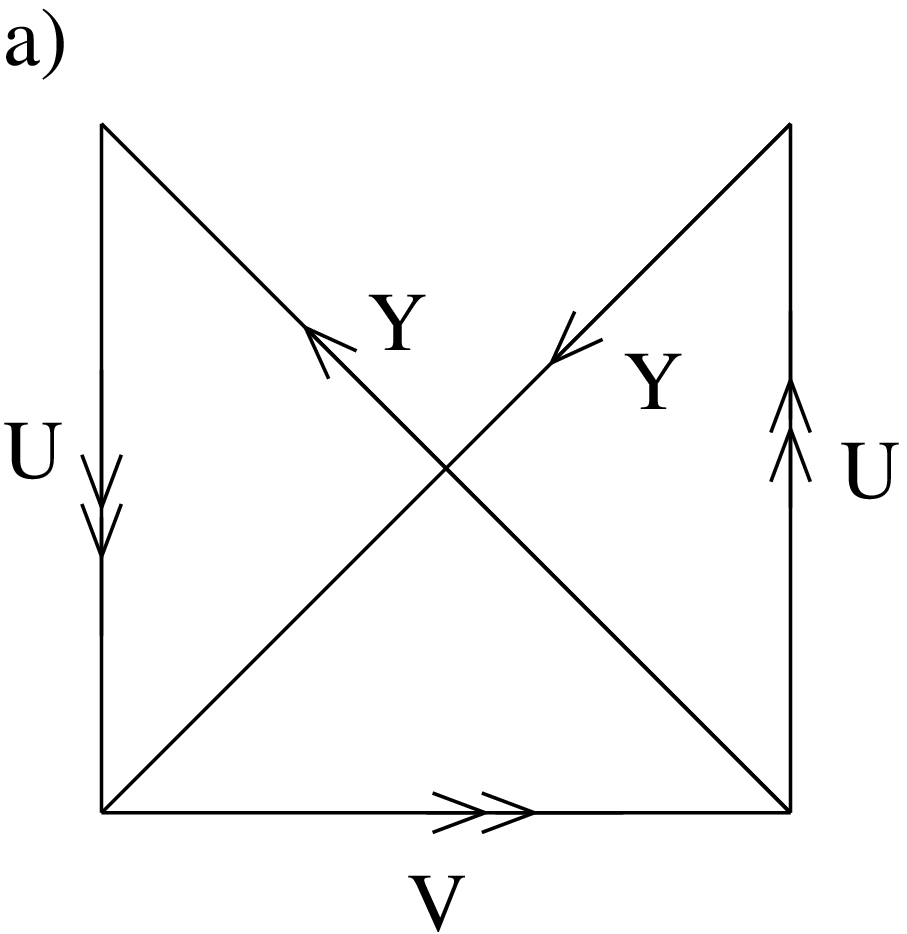}
\hfil
\includegraphics[width=2in]{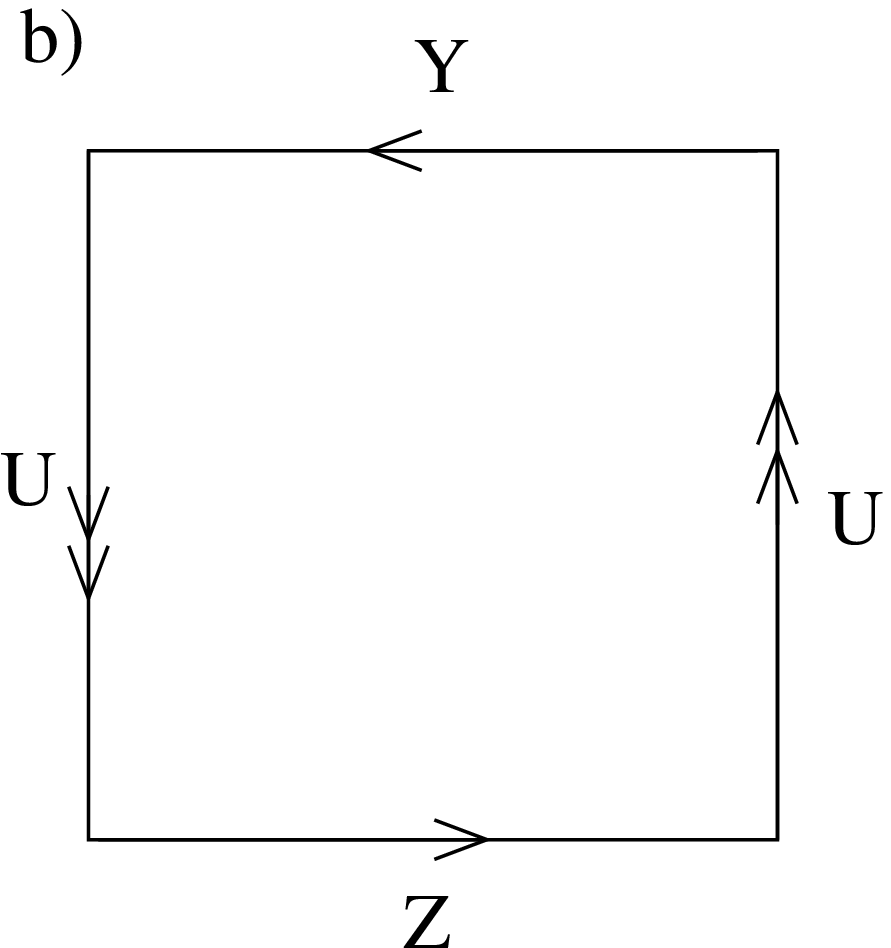} \\
\vfil
\includegraphics[width=4in]{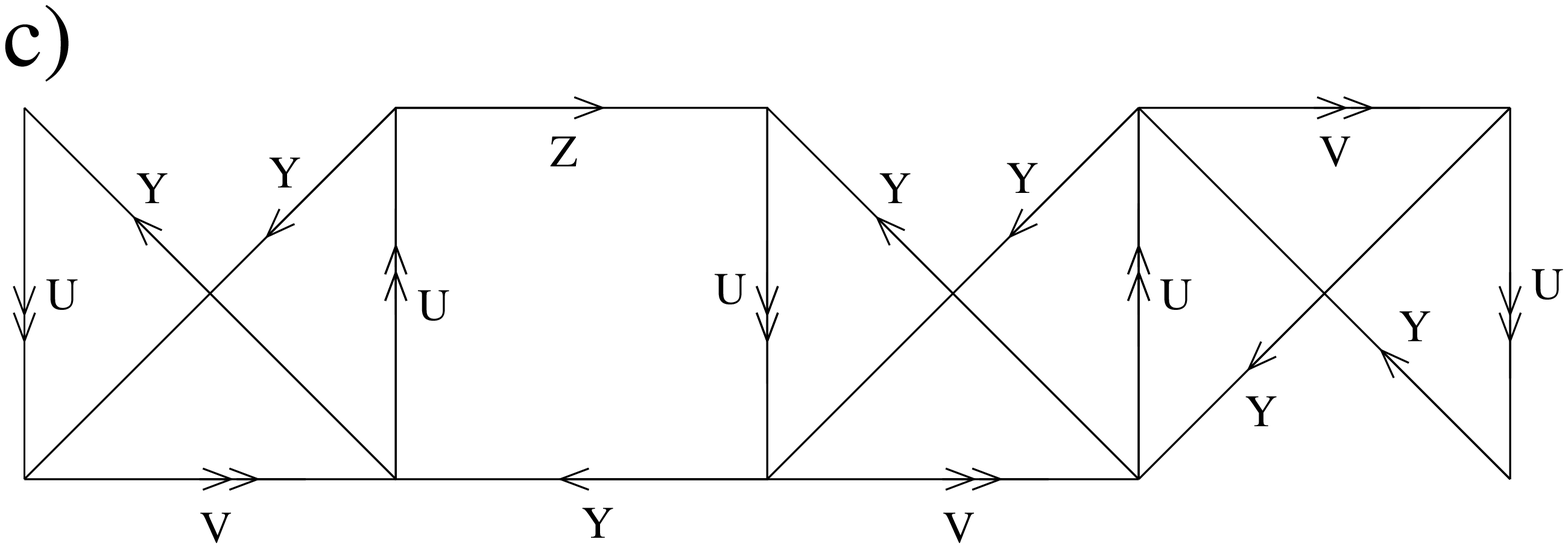}
\end{center}
\vfil
\caption{Shown are a) the unit cell $\sigma$; b) the unit cell $\tau$;
and c) the quiver for $Y^{4,3}$,
$\sigma \tilde \tau \sigma \tilde \sigma$.}
\label{unitcell}
\end{figure}

In this section, we review the construction of the $Y^{p,q}$ gauge
theories.
As derived in \cite{Benvenuti}, the quivers for these $Y^{p,q}$ gauge 
theories can be constructed from two basic units, $\sigma$ and $\tau$.
These units are shown in Figure \ref{unitcell}.
To construct a general quiver for $Y^{p,q}$, we define some basic
operations with $\sigma$ and $\tau$.  First, there are the inverted
unit cells, $\tilde \sigma$ and $\tilde \tau$, 
which are mirror images of $\sigma$ and
$\tau$ through a horizontal plane.  To glue the cells together,
we identify the double arrows corresponding to the $U^\alpha$
fields on two unit cells.  The arrows have to be pointing in 
the same direction for the identification to work.  So for
instance we may form the quiver $\sigma \tilde \tau = \tilde \tau \sigma$,
but $\sigma \tau$ is not allowed.  In this notation,
the first unit cell is to be glued not only to the cell on 
the right but also to the last cell in the chain.  A general 
quiver might
look like
\be
\sigma \tilde \sigma \sigma \tilde \tau \tau \tilde \sigma \ .
\ee
In general, a $Y^{p,q}$ quiver
consists of $p$ unit cells of which $q$ are of type $\sigma$.
The $Y^{p,p-1}$ gauge theories will have only one $\tau$ type 
unit cell, 
while the $Y^{p,1}$ theories will have
only one $\sigma$ type unit cell. 

Each node of the quiver corresponds to a gauge group while 
each arrow is a chiral field transforming in a bifundamental representation.
 For the $Y^{p,q}$ spaces, there are four types of bifundamentals
labeled $U^\alpha$, $V^\alpha$, $Y$, and $Z$ where $\alpha =1$ or 2.
To get a conformal theory, we take all the gauge groups to be $SU(N)$.

The superpotential for this quiver theory is constructed by summing
over gauge invariant operators cubic and quartic in the fields
$U^\alpha$, $V^\alpha$, $Y$, and $Z$.  For each $\sigma$ unit cell
in the gauge theory, we add two cubic terms to the superpotential
of the form
\be
 \epsilon_{\alpha \beta} U^{\alpha}_L V^\beta Y \; \; \mbox{and}
\; \;  \epsilon_{\alpha \beta} U^{\alpha}_R Y V^\beta \ .
\ee
Here, the indices $R$ and $L$ specify which group of $U^\alpha$ enter in the
superpotential, the $U^\alpha$ on the right side or the left side of $\sigma$.
The trace over the color indices has been suppressed.
 For each $\tau$ unit cell, we add the quartic term
\be
\epsilon_{\alpha\beta} U_L^\alpha Y U_R^\beta Z \ .
\ee
The signs should be chosen so that no phases appear in the relations.

\subsection{Chiral Primaries for General $Y^{p,q}$}

The quivers for general $Y^{p,q}$ are quite complicated, but some 
easy patterns emerge. 
We always expect to find the $b_1$, $b_2$, and $b_3$ type building blocks,
for these operators are just the independent superpotential type 
loops transforming in a
three dimensional representation of $SU(2)$.  At the superconformal point,
the R-charges of the $b_i$ are clearly all equal to two.
The $a_i$ and $c_i$ type operators also have analogs for general
$p$ and $q$.

For the $a_i$, it turns out there exists a loop in the quiver with
$p$ $Y$ type fields and $p-q$ type $U$ fields.  Such a loop naturally
transforms in a $p-q+1$ dimensional representation of $SU(2)$ and has
an R-charge
\be
R_a = p R_Y + (p-q) R_U \ .
\label{Raypq}
\ee

{}For the $c_i$, there is a loop with $p$ type $U$ operators,
$q$ type $V$ operators and $p-q$ type $Z$ operators.  Such a loop
transforms in a $p+q+1$ dimensional representation of $SU(2)$ and
has R-charge
\be
R_c = p R_U + q R_V + (p-q) R_Z \ .
\label{Rcypq}
\ee
Note that (\ref{Raypq}) and (\ref{Rcypq}) are consistent with the R-charges given
for $R_a$ and $R_c$ in footnote \ref{fn:Rcharges}.

\begin{figure}
\begin{center}
\includegraphics[width=5in]{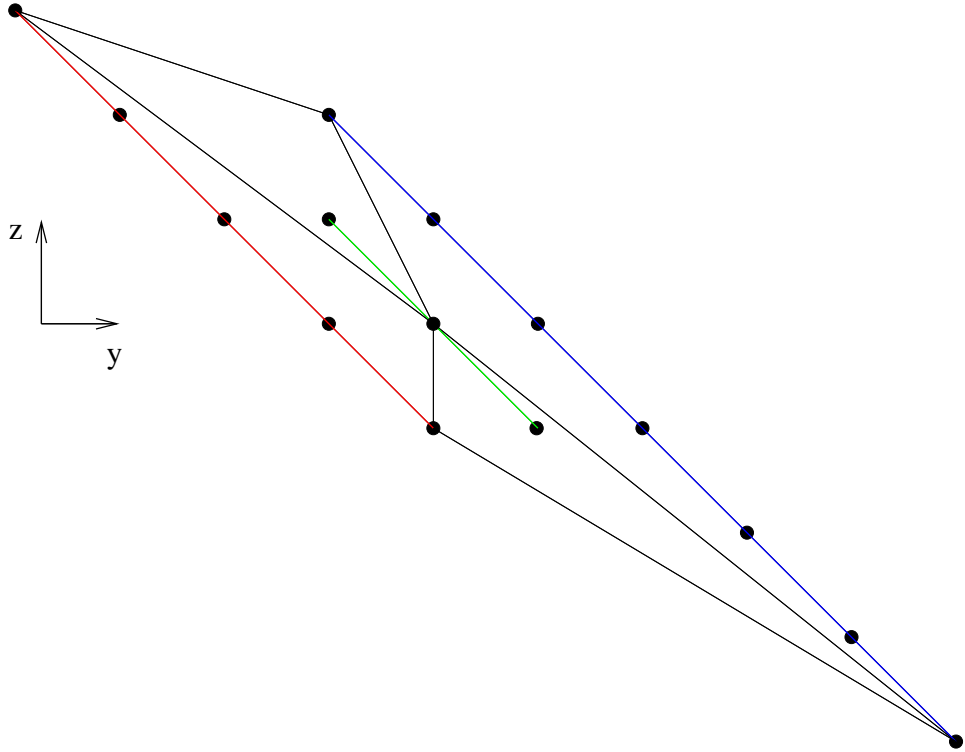}
\end{center}
\vfil
\caption{The dual cone $\sigma^\vee$ for $Y^{5,1}$.  The 
$a$ type operators correspond to lattice points along the red
line, the $b$ type operators to lattice points along the green
line, and the $c$ type operators to lattice points along the
blue line.}
\label{Y51}
\end{figure}

One can also see the $a$, $b$, and $c$ type operators emerge from the toric
diagram for the cones over the $Y^{p,q}$.  Recall \cite{Martelli}
that the toric cone is given
by the four vectors:
\be
e_1 = (1,0,0) \; ; \; \; \;
e_2 = (1,1,0) \; ; \; \; \;
e_3 = (1,p,p) \; ; \; \; \;
e_4 = (1, p-q-1, p-q) \ .
\ee
The dual cone $\sigma^\vee$ must then be bounded by the vectors orthogonal
to the faces of $\sigma$:
\be
e_1^\vee = (0,0,-1) \; ; \; \; \;
e_2^\vee = (-p,p,-p+1) \; ; \; \; \;
e_3^\vee = (-p, -q, q+1) \; ; \; \; \;
e_4^\vee = (0,-p+q, p-q-1) \ .
\ee

We now study $\sigma^\vee$.  Notice that $e_1^\vee$ and $e_4^\vee$ lie
along the line $x=0$ and $z=-y-1$.  This line will thus include $p-q+1$ lattice
points, and we tentatively identify these points with the $a_i$.  Similarly, 
$e_2^\vee$ and $e_3^\vee$ lie along the line $x=-p$ and $z=-y+1$.  This line
includes $p+q+1$ lattice points, and we identify these points with the $c_i$.

Finally, consider the three lattice points $(-1,0,0)$, $(-1,1,-1)$ and $(-1,-1,1)$.
These three points would be good candidates for the $b_i$ -- the points would
lead to expected relations of the form $ac \sim b^p$.  Note that
$(-1,1,-1)$ lies on the plane spanned by $e_1^\vee$ and $e_2^\vee$:
\be
(-1,1,-1) = \frac{1}{p} (e_1^\vee + e_2^\vee) \ ,
\ee
and similarly
\be
(-1,-1,1) = \frac{1}{p} (e_3^\vee + e_4^\vee) \ .
\ee
Thus these three lattice points do indeed lie inside $\sigma^\vee$.

These three sets of lattice points $a_i$, $b_i$, and $c_i$
span the cone $\sigma^\vee$.  More precisely, we expect that for any lattice point $p$
inside $\sigma^\vee$,
\be
p = \sum \ell_i a_i + \sum m_i b_i + \sum n_i c_i
\ee
for $\ell_i$, $m_i$ and $n_i$ non-negative integers.

\section{From the Metric to the Chiral Ring}
\label{sec:lap}

In this section, we will relate the $a_i$, $b_j$, and $c_k$ building blocks
of the chiral ring directly to the metric coordinates on the Calabi-Yau
cone.  
It turns out that it is just as easy to 
work out the relationship for all the $Y^{p,q}$ at once.

We choose coordinates on $Y^{p,q}$ such that the Sasaki-Einstein metric takes the form
\cite{Gauntlett2, Gauntlett}
\begin{eqnarray}
ds^2 &=& \frac{1-y}{6}\left( d\theta^2 + \sin^2 \theta d\phi^2 \right)+ \frac{1}{w(y)v(y)}dy^2 \nonumber \\
&+& \frac{v(y)}{9}\left( d\psi -\cos \theta d\phi\right)^2 
+ w(y) \left(d\alpha + f(y)(d\psi -\cos\theta d\phi) \right)^2
\end{eqnarray}
with the three functions $f(y),v(y), w(y)$ given by
\begin{eqnarray}
w(y) &=& 2\frac{b-y^2}{1-y} \\
v(y) &=& \frac{b-3y^2+2y^3}{b-y^2}\\
f(y) &=& \frac{b-2y+y^2}{6(b-y^2)}.
\end{eqnarray}

 For the metric to be complete,
\be
b = \frac{1}{2} - \frac{p^2 - 3q^2}{4p^3}\sqrt{4p^2-3q^2} \ .
\ee
The coordinate $y$ is allowed to range between the two smaller roots
of the cubic $b-3y^2+2y^3$:
\begin{eqnarray}
y_1 &=& \frac{1}{4p}
\left( 2p - 3q - \sqrt{4p^2-3q^2} \right) \ , \\
y_2 &=& \frac{1}{4p}
\left( 2p + 3q - \sqrt{4p^2-3q^2} \right) \ .
\end{eqnarray}
The period of $\alpha$ is $2\pi \ell$ where
\be
\ell = - \frac{q}{4 p^2 y_1 y_2} \ .
\ee
The remaining coordinates are allowed the following ranges:
$0 \leq \theta < \pi$, $0 \leq \phi < 2\pi$, and $0 \leq \psi < 2\pi$.

In these coordinates the scalar Laplacian is 
\begin{eqnarray}
\nabla^2 &=&
\frac{6}{1-y} \frac{1}{\sin\theta} \frac{\partial}{\partial \theta} \sin \theta \frac{\partial}{\partial\theta}
+ \frac{6}{1-y} \frac{1}{\sin^2\theta} \frac{\partial^2}{\partial \phi^2} \\
&& + \left(\frac{9}{v} - \frac{6}{1-y} + \frac{6}{1-y} \frac{1}{\sin^2\theta} \right)
\frac{\partial^2}{\partial \psi^2} + \frac{1}{1-y} \frac{\partial}{\partial y} vw (1-y) \frac{\partial}{\partial y}
\nonumber \\
&& +\left(\frac{9f^2}{v} + \frac{1}{w} \right) \frac{\partial^2}{\partial \alpha^2} 
+ \frac{12}{1-y} \frac{\cos\theta}{\sin^2\theta} \frac{\partial^2}{\partial \phi \partial \psi}
- \frac{18f}{v} \frac{\partial^2}{\partial \alpha \partial \psi} \ .\nonumber
\end{eqnarray}


%

\subsection{Chiral Primary Solutions}

The Laplacian on the cone $dr^2 + r^2 ds^2$ over $Y^{p,q}$ can be written
as
\be
\Box = \frac{1}{r^5} \frac{\partial}{\partial r} r^3 \frac{\partial}{\partial r} + \frac{1}{r^2} \nabla^2 \ .
\ee
This cone is a Kaehler manifold and $\Box = 2(\partial \partial^\dagger + \partial^\dagger \partial)$.
Thus any holomorphic function should satisfy the Laplace equation, $\Box \omega = 0$.
These holomorphic functions are our chiral primary operators.

Martelli and Sparks \cite{Martelli} provide three meromorphic functions on the cone over $Y^{p,q}$:
\begin{eqnarray}
z_1 &=& e^{i\phi} \tan \frac{\theta}{2}  \\
z_2 &=& e^{-6i\alpha - i \psi}  \frac{2}{\sin\theta} \prod_{i=1}^3 (y-y_i)^{-\frac{1}{2 y_i}} \\
z_3 &=& \frac{1}{2} r^3 e^{i\psi} \sin \theta \prod_{i=1}^3 (y-y_i)^{1/2} \ .
\end{eqnarray}
One can check that away from their singularities, these $z_i$ satisfy $\Box z_i = 0$.
Of these three functions, $z_3$ is also holomorphic.  The function
$z_1$ has a singularity at $\theta = \pi$, while $z_2$ has singularities at
$\theta = 0$, $\theta = \pi$, and $y=y_2$.  (Recall that $y_1 <0$.)
The function $z_2$ has the additional pathology of not being periodic under shifts
$\alpha \to \alpha + 2 \pi \ell$.

Using these three $z_i$, we would like to assemble a family of better behaved
holomorphic functions on the cone.  For this family, we assume the ansatz
\begin{eqnarray}
F_{mna}(z_i) &=& z_1^m z_2^{-a/6} z_3^{n-a/6} \\ \nonumber 
&=& 
r^{3n - a/2}
e^{im\phi} e^{in\psi} e^{ia\alpha} \sin^{n+m} \frac{\theta}{2} 
\cos^{n-m} \frac{\theta}{2}  \prod_{i=1}^3 (y-y_i)^{e_i}
\end{eqnarray}
where 
\be
e_i = \frac{1}{12y_i} \left( a(1-y_i) + 6ny_i \right) \ .
\ee
To be free of singularities in $\theta$, we must take $-n \leq m \leq n$.
To be free of singularities in $y$,
we have the two conditions 
$e_i \geq 0$
for $i=1$, 2.
We also have a number of periodicity constraints.
For periodicity in $\alpha$, we need $a = P/\ell$ where $P$ is an integer.
We will assume that $\phi$ is periodic under shifts by $2\pi$ and that
$\phi + \psi$ is periodic under shifts by $4\pi$.
These constraints then imply that $m$ and $n$ are either
both integer or both half-integer.

One easy set of holomorphic functions can be found by setting $a=0$.
We find
\begin{eqnarray}
b_1 &=& z_3 z_1^{-1} = r^3 e^{i(\psi -\phi)} \cos^2 \frac{\theta}{2} \prod_{i=1}^3 (y-y_i)^{1/2} \\
b_2 &=& z_3 = r^3 e^{i\psi} \sin \frac{\theta}{2} \cos \frac{\theta}{2} \prod_{i=1}^3 (y-y_i)^{1/2} \\
b_3 &=& z_3 z_1 = r^3 e^{i(\psi +  \phi)} \sin^2 \frac{\theta}{2} \prod_{i=1}^3 (y-y_i)^{1/2}
\end{eqnarray}
where we have identified these objects with the holomorphic monomials of
the previous section.  Superficially, $b_1$ seems not to be single valued at $\theta=0$
while $b_3$ would not be single valued at $\theta=\pi$.  In fact, at these poles,
the metric on $Y^{p,q}$ degenerates such that the good angular coordinate 
at $\theta = 0$ is no longer $\psi$ but $\psi-\phi$ and at $\theta=\pi$, $\psi+\phi$.

To find the other holomorphic functions, 
we explore the boundaries of the singularity conditions on $y$.
In particular, consider the limit $e_i=0$.
In the case $i=1$, we find that 
\be
a = -\frac{6n y_1}{1-y_1} \; ; \; \; \; e_2 = \frac{np}{q+p} \ .
\ee
The periodicity condition on $\alpha$ implies that
\be
n = \frac{q+p}{2} P \ .
\ee
Choosing the smallest nontrivial value $P=1$ that avoids
a singularity at $y=y_2$, we find that
$e_2 = p/2$.
These operators take the form
\be
c_{i_m} = z_1^m r^\Delta e^{i (p+q) \psi / 2 + i \alpha / \ell} 
\left(\sin \frac{\theta}{2} \cos \frac{\theta}{2} \right)^{(p+q)/2}
(y-y_2)^{p/2} (y-y_3)^{e_3} \ ,
\ee
where the R-charge 
\be
R = \frac{2}{3} \Delta = p \frac{2y_2}{y_2-y_1} \ ,
\label{Rcmetric}
\ee
and
\be
e_3 = \frac{q+p}{4} \left(1 - \frac{y_1(1-y_3)}{y_3(1-y_1)} \right) \ .
\ee
We have labelled these operators to suggest the relationship to the $c_i$ 
of the previous section.  These operators fill out a $q+p+1$ dimensional representation
of $SU(2)$.

Next consider the limit $e_2=0$, in which case
\be
a = -\frac{6n y_2}{1-y_2} \; ; \; \; \; e_1 = \frac{np}{p-q} \ .
\ee
Periodicity on $\alpha$ now implies that
\be
n = -\frac{p-q}{2} P \ .
\ee
Choosing the smallest nontrivial value of $P=-1$ that avoids
a singularity, we find that $e_1 = p/2$.  These operators
take the form
\be
a_{i_m} = z_1^m r^\Delta e^{i (p-q) \psi / 2 - i \alpha / \ell} 
\left(\sin \frac{\theta}{2} \cos \frac{\theta}{2} \right)^{(p-q)/2}
(y-y_1)^{p/2} (y-y_3)^{e_3} \ ,
\ee
where the R-charge 
\be
R = \frac{2}{3} \Delta = -p \frac{2y_1}{y_2-y_1} \ ,
\label{Rametric}
\ee
and
\be
e_3 = \frac{p-q}{4} \left(1 - \frac{y_2(1-y_3)}{y_3(1-y_2)} \right) \ .
\ee
These operators fill out a $p-q+1$ dimensional representation of $SU(2)$ like
the $a_i$ of the previous section.

Using for example the table in \cite{Benvenuti}, one may easily check that the R-charges
computed from these holomorphic functions agrees with the R-charges computed
from field theory.  In other words, 
(\ref{Rametric}) agrees with (\ref{Raypq}) and (\ref{Rcmetric}) is eqal to 
(\ref{Rcypq}).

We now check that these $a_j$ and $c_j$ are indeed single valued at $y=y_2$ and
$y=y_1$ respectively.  From the metric, we see that at $y=y_i$, the good angular
coordinate is no longer $\alpha$ but
\be
\alpha + f(y_i) \psi = \alpha + \frac{1}{6} \frac{y_i-1}{y_i} \psi\ .
\ee
Correspondingly, the exponent for the $a_j$ can be written
\be
\frac{p-q}{2} \psi - \frac{1}{\ell} \alpha = -\frac{1}{\ell} \left( \alpha + \frac{(q-p)\ell}{2} \psi \right)
\ee
and it is not difficult to check that
\be
\frac{(q-p) \ell}{2} = \frac{y_2-1}{6y_2} \ .
\ee
For the $c_j$, we find that
\be
\frac{p+q}{2} \psi + \frac{1}{\ell} \alpha = \frac{1}{\ell} \left( \alpha + \frac{(p+q)\ell}{2} \psi \right)
\ee
and
\be
\frac{(p+q)\ell}{2} = \frac{y_1-1}{6y_1} \ .
\ee

\end{document}